\documentstyle[epsfig,pstcol]{aipproc}

\begin{document}
\title{         
Muon Dynamics in a Toroidal Sector Magnet}
\author{Juan C. Gallardo\thanks{Email:gallardo@bnl.gov}, Richard C. Fernow and Robert B. Palmer} 
\address{Physics Department, 
 Brookhaven National Laboratory, Upton, NY 11973, 
USA\\}
\maketitle

\begin{abstract} 
We present a Hamiltonian formulation of muon dynamics in  toroidal sector solenoids (bent solenoid). 
\end{abstract}

\section{INTRODUCTION}

The present scenario for the cooling channel in a high brightness muon 
collider\cite{ref1} calls for a quasi-continuous solenoidal focusing channel. 
The beam line consists of a periodic array of  hydrogen absorbers 
 immersed in a solenoid  with alternating focusing field and rf linacs at the zero field points.

 The simple ${dE\over dx}$ energy loss in conjunction with multiple scattering 
 and energy straggling leads to a decrease of the normalized transverse 
 emittance. Reduction of the longitudinal emittance could be achieved by wedges 
 of material located in dispersive regions; at least in principle, this scenario 
 seems appropriate to obtain effective 6-D phase space reduction.\cite{ref2}

A conventional chicane is a dispersion element but its use presents a 
serious challenge, as it is very difficult to integrate it with the solenoidal  
channel. The matching into the periodic solenoidal system imposes constraints on the Twiss parameters of the beam which seems not easily achievable. A possible alternative is the use of curved solenoids in conjunction with wedge absorbers as suggested by one of the authors.\cite{ref3} 

\begin{figure}
\begin{center} 
\newpsobject{showgrid}{psgrid}{subgriddiv=1,griddots=10,gridlabels=6pt}
\begin{pspicture}(5,0)(13,9)
\psline[linecolor=blue](5,1)(6,1)
\psline[linecolor=blue](5,2)(6,2)
\psdots[linecolor=blue,dotstyle={square*},dotscale=2](5,1)
\psdots[linecolor=blue,dotstyle={square*},dotscale=2](5,2)
\psarc[linecolor=green](6,4){2}{270}{360}
\psarc[linecolor=green](6,4){3}{270}{360}
\psline(8,4)(8,6)
\psline(9,4)(9,6)
\pswedge[fillstyle=vlines,linecolor=blue,
hatchangle=20](7,5){3}{-2.5}{2.5}
\uput[ur](9.5,5.25){Wedge absorber}
\psarc[linecolor=green](11,6){2}{90}{180}
\psline[linecolor=red]{->}(6,4)(7.520489,2.700726)
\uput[ur](7.1,3.1){$R_0$}
\psarc[linecolor=green](11,6){3}{90}{180}
\psline[linecolor=red]{->}(8.28073,2.0511)(8.6608,1.7262)
\uput[ur](8.6608,1.6262){$\vec{e}_x$}
\psline[linecolor=red]{->}(8.28073,2.0511)(8.6608,2.37592)
\uput[ur](8.6608,2.27592){$\vec{e}_s$}
\psline[linecolor=blue](11,8)(13,8)
\psline[linecolor=blue](11,9)(13,9)
\psdots[linecolor=blue,dotstyle={square*},dotscale=2](13,8)
\psdots[linecolor=blue,dotstyle={square*},dotscale=2](13,9)
\end{pspicture}     

\caption{Schematic of emittance exchange system with a bent solenoid}\label{fig1}
\end{center}
\end{figure}
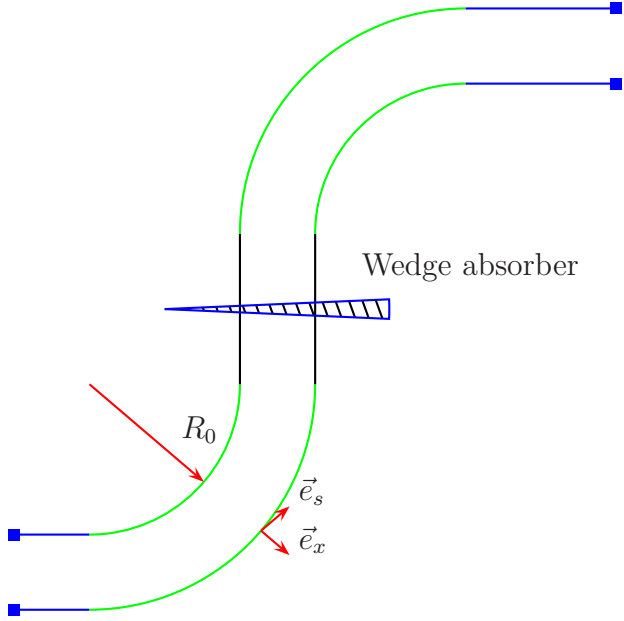

Solenoids and toroidal sectors have a natural place in muon collider design 
given the large emittance of the beam and consequently, the large transverse  momentum  of the initial pion beam or the decay muon beam. Bent solenoids as  shown in Fig.\ref{fig1} 
were studied for use at the front end of the machine, as part of the capture 
channel\cite{ref4} and more recently as part of a diagnostic setup to 
measure the position and momentum of muons.\cite{ref5}
\section{Toroidal Sector Solenoid}                                                                           
If we restrict ourselves for the moment to a horizontal bending plane, the magnetic 
field inside of the solenoid and near the axis has a gradient (field lines are denser at smaller 
radius) described approximately by $\vec{B}(x,y,s)\approx B_s\vec{e}_s$ with
\begin{equation}
B_s(x,y,s)\approx {B_s(0,0,s)\over (1+hx)}\label{eq1}
\end{equation}
where $s $ is the coordinate along the particle trajectory and 
$h={1\over R_o} $ is the curvature at the position s, with $R_o$ the 
radius of curvature. As a consequence of the curvature of the trajectory 
and the corresponding magnetic gradient, the center of the particle 
guide orbit, averaged over the Larmor period, drifts in a direction 
perpendicular to the plane of bending\cite{ref6}. The combined drift velocity can 
be written as,
\begin{equation}
{d\vec{r}\over dt}=v_{\|}{\vec{B}\over B}+{m_{\mu}\over 2 q_{\mu}
}(2v_{\|}^2+v_{\perp}^2){(\vec{R_o}\times \vec{B})\over (1+hx)R_o^2B^2}.\label{eq2}
\end{equation}
and the magnitude of the transverse drift velocity is 
\begin{equation}
v_{\rm drift}^T ={m_{\mu}\over 2 q_{\mu}}
(2v_{\|}^2+v_{\perp}^2){h\over (1+hx)B_s}
\end{equation}
Clearly a y-position versus energy ($v_{||}$) correlation will develop 
as the muon beam travels along the toroidal sector solenoid.

From Eq.\ref{eq2} above we notice that if we include an additional 
vertical field, a dipole with a curvature equal to that of the 
bent solenoid for the reference energy, i.e.
 \begin{equation}
\vec{B}_D \approx -{|B|\over v_{||}}v_{\rm drift}^T\vec{e}_y
 \end{equation}
then Eq.\ref{eq2} reduces to,
 \begin{equation}
{d\vec{r}\over dt}= v_{||}{B_s\over |B|}\vec{e}_s+(v_{||}-v_{||}^o){B_D\over |B|}\vec{e}_y
 \end{equation}
and consequently, particles with the chosen energy will not drift 
vertically and will remain on the axis of the bent solenoid. 
Those particles with larger energy will drift upward (positive y-direction) 
and those with lower energy downward (negative y-direction), achieving the needed dispersion. The magnitude of the dispersion is given by 
\begin{equation}
D_y= 2\pi {p_o \over q}{B_D\over B_s^2} = 2\pi h \left({p_o\over q B_s}\right)^2
\label{dispersion}
\end{equation}
where $p_o$ is the chosen momentum corresponding to zero dispersion.

\section{Dynamics in a Toroidal Solenoid. Hamiltonian Formulation}
From general principles of classical mechanics and following the usual approximations 
in accelerator physics\cite{ref7} the normalized Hamiltonian reads,
\begin{eqnarray}
&H(x,p_x,y,p_y,z,\delta;s)\approx  -{q\over p_0}(1+hx)A_s(s)-(1+\delta)hx \nonumber \\
&+{(1+hx)\over 2(1+\delta)}
\left\{ (p_x-{q\over p_0}A_x)^2+
(p_y-{q\over p_0}A_y)^2\right\}
\label{eq5}
\end{eqnarray}
where the path length $s$ is the independent variable, $p_x\,,\, p_y$ are normalized 
momenta with respect to $p_0$, the initial reference total momentum $p_0=
\sqrt{p_x^2+p_y^2+p_s^2}$; $z=s-\beta_o c t$, $\delta={(p-p_0)\over p_0}$ and 
$\vec{A}=(A_x, A_y, A_s)$ is the vector potential. The vector potential 
satisfies the gauge condition $\nabla \cdot \vec{A}=0.$

In the accelerator frame of reference, i.e. the Frenet-Serret coordinate 
system defined by the metric,
\begin{equation}
     d\sigma^2=dx^2 +dy^2 +(1+hx)^2 ds^2
     \label{eq6}
\end{equation}
the equations for the coordinate unit vectors are
\begin{equation}
{d\vec{e_x}\over ds}=h(s) \vec{e_s}\quad , \quad 
{d\vec{e_y}\over ds}=0 \quad , \quad   
{d\vec{e_s}\over ds}=-h(s)\vec{e_x} 
\end{equation}
The magnetic and electric fields are obtained from
\begin{equation}
\vec{B}=\nabla \times \vec{A} \qquad {\rm and} \qquad 
\vec{E} =c\beta_0 {\partial\vec{A}\over \partial z} 
\end{equation}
with 
\begin{eqnarray}
\nabla \times \vec{A} =&{1\over (1+hx)} \left\{ \partial_y (1+hx)A_s -\partial_sA_y\right\}\vec{e}_x \nonumber \\
&+{1\over (1+hx)}\left\{\partial_s A_x -\partial_x (1+hx) A_s \right\}\vec{e}_y \nonumber \\
&+\left\{ \partial_xA_y-\partial_yA_x\right\}\vec{e}_s
\end{eqnarray}
and
\begin{equation}
\nabla \cdot \vec{A}={1\over (1+hx)}\left[ {\partial \over \partial x} \left( A_x(1+hx)\right) + 
{\partial \over \partial y} \left( A_y(1+hx)\right) +{\partial \over \partial s}A_s\right]
\end{equation}

\begin{figure}[th]
\begin{center}
\includegraphics[height=4in,width=5.0in,angle=90]{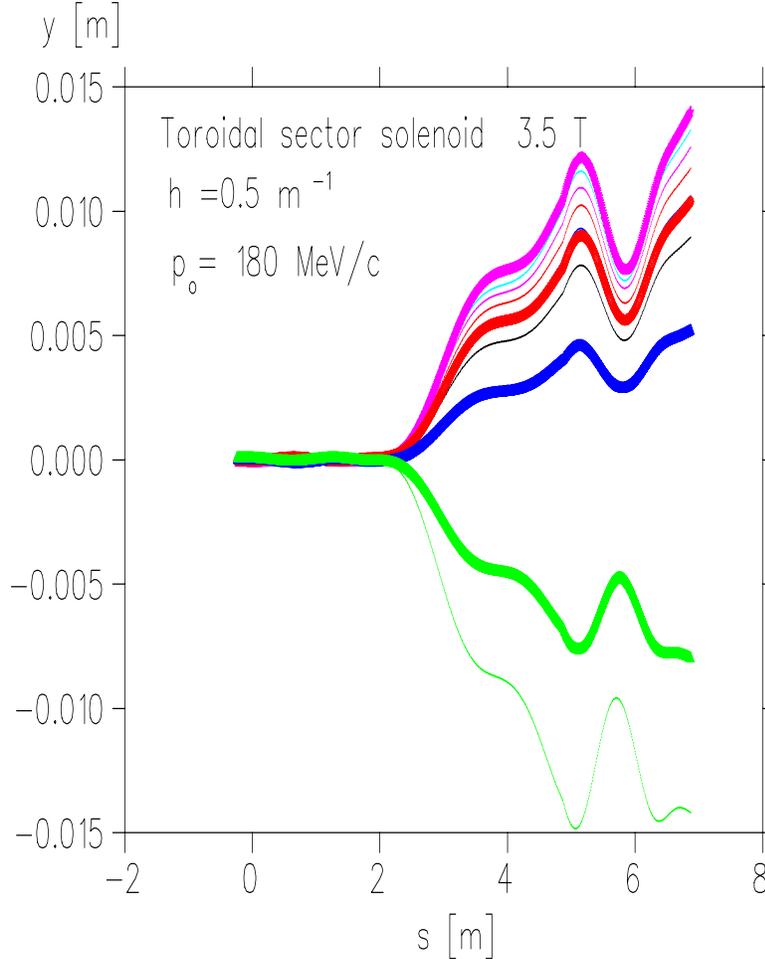}
\caption{Example of dispersion created by a toroidal sector solenoid plus a vertical dipole. We show several tracks with energies ($\pm 1\%$) larger and smaller than the reference energy $p_o.$}\label{fig2}
\end{center}
\end{figure}

The lowest order approximation for a toroidal solenoidal field
 is given by Eq.\ref{eq1}. The corresponding vector potential in 
 the next order is\cite{ref8},
 \begin{equation}
\vec{A} = -{1\over 2} B_o {y\over (1+hx)}\vec{e}_x +{B_o\over 2 h}\ln{(1+hx)}\vec{e}_y
\end{equation}
and the corresponding magnetic fields are:
\begin{mathletters}
\label{field}
\begin{eqnarray}
&B_x=-{1\over (1+h(s)x)}{1\over 2h(s)}\left\{ {dB_o\over ds}\ln{(1+h(s)x)}\right. \nonumber \\
&-\left.{B_o\over h(s)}h'(s)\ln{(1+h(s)x)} +h'(s){B_ox\over (1+h(s)x)}\right\}
\end{eqnarray}
\begin{equation}
B_y=-{1\over 2 (1+h(s)x)^2}\left\{{dB_o\over ds}y -h'(s){B_o(s)xy\over 1+h(s)x}\right\}
\end{equation}
\begin{equation}
B_s= {B_o\over 1+h(s)x}
\end{equation}
\end{mathletters}
 One possible  second order approximation for the  vector potential of a dipole is
\begin{mathletters}
 \begin{eqnarray}
A_x& =& -{B_D\over 2 h(s)}{h''(s)\,x(y^2-{1\over 3}x^2)\over (1+h(s)x)} \\
A_y &= &{B_D\over 2 h(s)}{h'(s)\,xy\over (1+h(s)x)}\\
A_s &=& -{B_D\over 2 h(s)}\left\{1+h(s)\,x-h'(s)\,(y^2-x^2)\right\} 
 \end{eqnarray}
\end{mathletters}
Substituting the total vector potential into the Hamiltonian, and dropping some constants we can  write,
\begin{eqnarray}
&H_{\rm tor. sol.}^{\rm dip}(x,p_x,y,p_y,z,\delta;s)\approx {1\over 2} (hx)^2-\delta hx \nonumber \\
&+{(1+hx)\over 2(1+\delta)}\left\{(p_x-{q\over p_o}A_x)^2+(p_y-{q\over p_o}A_y)^2\right\}
\end{eqnarray}
We have written a  simple Fortran program to solve the  equations of motion 
from the above Hamiltonian ; its result for a few representative cases of 
interest are shown in Fig.\ref{fig2}.

\begin{figure}[th]
\begin{center}
\includegraphics[height=4in,width=4.0in]{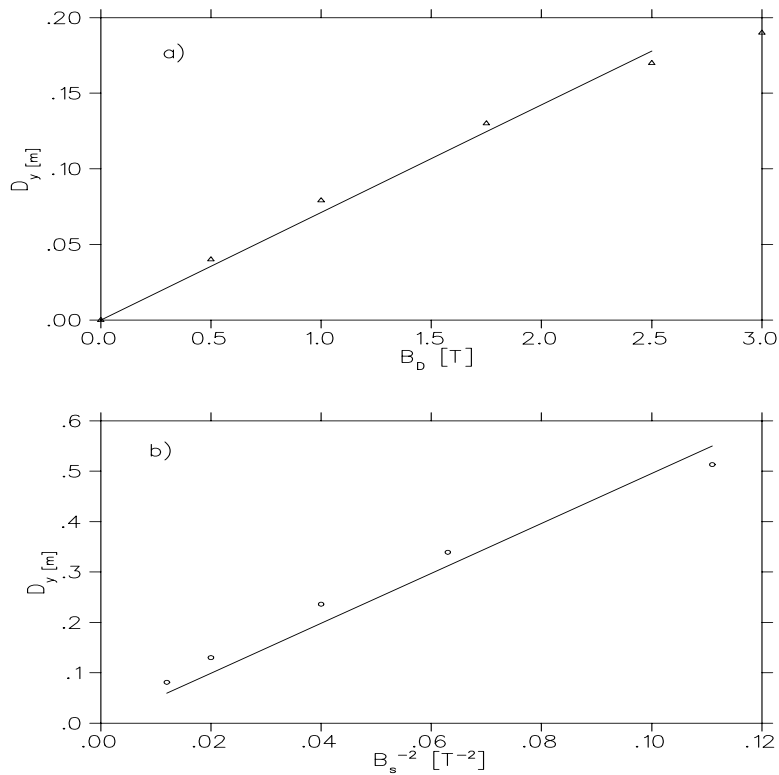}
\caption{a) Dispersion, $D_y$ vs. dipole magnetic field, $B_D$; 
b) Dispersion, $D_y$ vs. solenoid magnetic field $B_s^{-2}.$ }\label{sffg2}
\end{center}
\end{figure}

 A second order expansion of the bent solenoid magnetic field given in 
Eqs. \ref{field} has been used together with a second order expansion of the 
dipole magnetic field in the cooling simulation program ICOOL\cite{ref9}. 
Fig.\ref{sffg2}a shows an example of the dispersion $D_y$ in a bent solenoid 
obtained in ICOOL as a function of the dipole strength $B_D.$ It is apparent 
that the dependence of Eq. \ref{dispersion} on $B_D$ is well satisfied. Likewise, Fig \ref{sffg2}b shows simulation results for the dispersion as a function of $B_s^{-2}.$ Again we see that the mentioned equation gives a good representation of the results.  
\section{Acknowledgements } 
This work was supported by the US DoE under Contract 
No. DE-AC02-76CH00016.


\begin{thebibliography}{8}
\bibitem{ref1} R. B. Palmer, presentation at FNAL Muon 
Collider Collaboration Meeting, July 1995, unpublished;  
R. B. Palmer, et al., {\it Muon Colliders}, 9th Advanced 
ICFA Beam Dynamics Workshop, Ed. Juan C. Gallardo, AIP Press, 
AIP Conference Proceedings 372, (1996).
\bibitem{ref2} R. B. Palmer, {\it Cooling Theory}, in preparation.
\bibitem{ref3} R. B. Palmer, private communication.
\bibitem{ref4}\textbf{ $\mathbf{\mu^+\mu^-}$ Collider: A Feasibility Study}, New Directions for High-Energy Physics. Proceedings of the 1996 DPF/DPB Summer Study on High-Energy Physics Snowmass'96, Chapter 4; see also the Muon Collider 
Collaboration WEB page {\it http://www.cap.bnl.gov/mumu/}
\bibitem{ref5} C. Lu, K. T. McDonald and E. J. Prebys, {\it A Detector Scenario for the Muon Cooling Experiment}, Princeton/$\mu \mu$/97-8, July 1997.
\bibitem{ref6} F. Chen, {\it Introduction to Plasma Physics}, Chap.2, Plenum Press (1979). 
\bibitem{ref7} C. Wang and A. Chao, {\it Notes on Lie algebraic analysis of achromats}, SLAC/AP-100, Jan. 1995; E. D. Courant and H. S. Snyder, {\it Theory of the alternating-gradient synchrotron}, Ann. of Phys. 3,1 (1958); R. Ruth, {\it Single Particle Dynamics in Circular Accelerators}, Physics of Particle Accelerators, AIP Conference Proc. 153, Ed. M. Month and M. Dienes, Vol.1, pag. 150, (1987); C. Wang and A. Chao, {\it Transfer matrices of superimposed magnets and RF cavity}, SLAC/AP-106, Nov. 1996; H. Wiedemann, {\it Particle Accelerator Physics II}, pag. 51, Springer (1995).
\bibitem{ref8} See C. Wang and A. Chao\cite{ref7} and A. Morozov and Solov'ev, {\it Motion of charged particles in e.m. fields}, Review of Plasma Physics, vol. II, Ed. M. A. Leontovich, Consultants Bureau, Division of Plenum Publishing Company, New York (1966).
\bibitem{ref9} R. Fernow, ICOOL, fortran program to simulate muon ionization cooling.

\end{thebibliography}
\end{document}